# Intelligent Active Queue Management Using Explicit Congestion Notification

Cesar A. Gomez, Xianbin Wang, and Abdallah Shami
Department of Electrical and Computer Engineering, Western University, London, Ontario, Canada
Email:{ cgomezsu, xianbin.wang, abdallah.shami }@uwo.ca

*Abstract*—As more end devices are getting connected, the Internet will become more congested. Various congestion control techniques have been developed either on transport or network layers. Active Queue Management (AQM) is a paradigm that aims to mitigate the congestion on the network layer through active buffer control to avoid overflow. However, finding the right parameters for an AQM scheme is challenging, due to the complexity and dynamics of the networks. On the other hand, the Explicit Congestion Notification (ECN) mechanism is a solution that makes visible incipient congestion on the network layer to the transport layer. In this work, we propose to exploit the ECN information to improve AQM algorithms by applying Machine Learning techniques. Our intelligent method uses an artificial neural network to predict congestion and an AQM parameter tuner based on reinforcement learning. The evaluation results show that our solution can enhance the performance of deployed AQM, using the existing TCP congestion control mechanisms.

*Keywords—Active Queue Management (AQM), congestion control, Explicit Congestion Notification (ECN), Machine Learning*

## I. INTRODUCTION

Thanks to the proliferation of smart devices and the paradigm of Internet of Things (IoT), the demand for connections to the Internet is dramatically growing. As a response, Internet Service Providers (ISPs) are focused on improving the performance of their networks and connections to the Internet. However, engineers and researchers are trying to address this challenge by solving the traditional networks' congestion problems. On the one hand, congestion avoidance mechanisms in TCP have been part of the solution and essential for the massive adoption of the World Wide Web. On the other hand, due to the bottlenecks along the paths, buffers have been deployed to avoid packet loss when packets arrive at faster rate than can the links. Nevertheless, excessive buffering leads to increasing delays, as packets have to stay longer in the queues, and causing a phenomenon known as bufferbloat [1]. Network devices tackle this effect through Active Queue Management (AQM) techniques, which aim to avoid the buffer's overflow by dropping or marking the packets before the buffer fills completely. A variety of AQM schemes has been proposed, including the classical Random Early Detection (RED) algorithm [2], the Controlling Queue Delay (CoDel) [3], and newer ones such as the Proportional Integral controller Enhanced (PIE) [4] and the Flow Queue CoDel (FQ-CoDel) [5]. Despite the advantages of AQM techniques, they are not widely adopted in ISPs' network devices for the following reasons:

first, some AQM mechanisms have parameters that might be difficult to tune in very dynamic environments. Second, routers and switches with more memory available in the market have created the misconception that the larger the buffers, the better.

The main advantage of dropping packets with AQM rather than with tail-drop queues, *i.e.* buffers with no AQM, is to avoid the unnecessary global synchronization of flows when a queue overflows. Consequently, network devices drop more packets when no AQM scheme is in use and the network throughput is deteriorated. In contrast, an AQM method can decide to either drop or mark packets when the network experiences incipient congestion. The process of marking packets instead of dropping them is known as Explicit Congestion Notification (ECN). The employment of ECN can reduce the packet loss and latency of Internet connections, among other benefits such as improving throughput, reducing probability of retransmission timeout expiry, and reducing the head-of-line blocking [6]. Moreover, the importance of ECN relies on its fact of making incipient congestion visible, by exposing the presence of congestion on a path to network and transport layers. The data containing ECN-marked packets can be exploited to learn some characteristics such as the level of congestion of a network operator and the behavior of TCP protocols or applications, for instance. For these reasons, the deployment of new ECN-capable end systems and the necessity of reducing queuing delay in modern networks have motivated the interest in ECN [7]. Indeed, IETF has published a significant number of RFC documents regarding ECN, which indicates strong level of interests from industry and academia.

ECN is specified in the RFC3168 [8], which defines four codepoints through two bits in the IP header, to indicate whether a transport protocol supports ECN and if there is congestion experienced (CE). This IETF recommendation also specifies two flags in the TCP header to signal ECN: the ECN-Echo (ECE) and the Congestion Window Reduced (CWR). Then, if the AQM algorithm in any router along the path determines that there is congestion, the router marks the packets with the CE code to indicate to the receiver that the network has experienced congestion. Once the CE-marked packet arrives at the receiver, it echoes back a packet to the sender with the ECE flag set in the TCP header to notify that congestion was experienced along the path. Consequently, the sender reduces the data transmission rate and sends the next TCP segment to the receiver with the CWR flag set. It is important to highlight that TCP also responds to non-explicit congestion indication produced by tail-drop queues or AQM dropping. How TCP performs those actions depends on the congestion control mechanisms on the transport layer and their details are out of the scope of this paper. However, it is evident that the utilization of ECN mitigates the need for packet retransmission and, consequently, avoids the excessive delays due to retransmissions after packet losses. In addition, without ECN it is not possible to determine if the packets are lost because of congestion or poor link quality. Finally, we point out the rest-of-path congestion concept introduced in the Congestion Exposure (ConEx) mechanism, which to some extent has inspired our work. Although proposed several years ago, the implementation of ConEx is not widely deployed, as it needs modifications to the TCP protocol at the sender side [9].

Accordingly, in this work we propose an intelligent use of the standardized ECN mechanism for existing AQM solutions. We build our method on Machine Learning techniques for the exploitation of ECN. The method consists of two main parts: a congestion predictor and a dynamic parameter tuner. The latter applies a Reinforcement Learning (RL) technique to balance the delay and throughput by adaptively setting the AQM parameters. The congestion predictor is a Neural Network (NN) that forecasts if there will be congestion on the rest-of-path. Our main goal is to propose a scheme that is fully compatible with existing TCP congestion control mechanisms and already deployed AQM techniques. Although previous works have used Machine Learning techniques to solve problems regarding AQM, to the best of our knowledge, none of them exploits ECN to improve the AQM mechanisms. For example, authors in [10] compare several AQM techniques based on NN with conventional AQM techniques. Through simulations, the authors show that the studied NN-based methods converge faster than the traditional techniques. Similarly, Bisoy and Pattnaik propose in [11] an AQM controller based on feed-forward NN, which stabilizes the queue length by learning the traffic patterns. Also, on the basis of RL, Bouacida and Shihada present in [12] the LearnQueue method, which focuses on the operation in wireless networks. Authors model their solution by adapting the Q-learning algorithm to control the buffer size. By means of unsupervised learning techniques, authors in [13] propose a cognitive algorithm to detect and penalize misbehaving ECN-enabled connections. Although this problem and the employed techniques differ from ours, we find some similarities in terms of exploiting the TCP connection data and the implementation on top of existing AQM mechanisms.

## II. INTELLIGENT AQM DESIGN

As we mentioned in the Introduction, our goal is to enhance the performance that current AQM techniques provide at bottlenecks. We have explained how the ECN can reduce the connections' latency when enabled in the AQM along a path. However, ECN is not currently exploited to estimate the congestion ahead and dynamically adjust the AQM parameters in a router. Our hypothesis is that TCP connections can have a better performance if AQM schemes are tuned based on the specific network conditions. Yet, this is a non-trivial problem due to the complexity of IP networks. Consequently, we propose an intelligent method for improving existing AQM that learns from the experience and ECN feedback of a changing network. Our method is meant to be implemented on edge routers for two main reasons: first, edge routers are more prone to experience congestion than core routers, due to the bottleneck link between the access network and the backbone. Second, our mechanism uses traffic data in the downstream direction, which may take different paths in the core network. Despite these reasons, our solution can be deployed in core network devices even if ECN feedback is not completely obtained. The overall scenario for our stated problem is shown in Fig. 1, which is a valid topology for end points connected through a shared bottleneck link [14]. It is also important to highlight that ECN is not a perfect mechanism for congestion control. If an AQM decides to mark every packet with incipient congestion regardless the status of the queue, the AQM could produce a harmful effect. That is why we argue that a right and dynamic setting of the AQM parameters is pertinent. Moreover, we point out the potential application of Machine Learning techniques for this purpose.

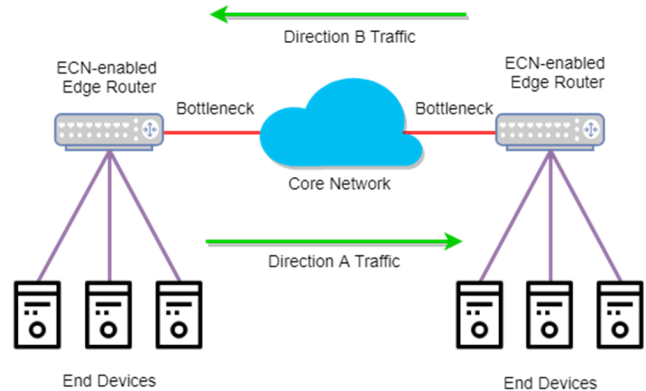

Fig. 1. Scenario for our stated problem. Edge routers aggregate end devices and connect to the core network through bottleneck links.

### A. Congestion Predictor

To predict the congestion, we take advantage of the ECE flag available in the TCP header of the packets in direction B without considering the ones involved in the ECN negotiation, as those packets indicate the setting of ECN-capable TCP sessions rather than congestion or response to congestion [8]. We model the congestion prediction as a time-series problem. The core of the congestion predictor is a Long Short-Term Memory (LSTM), which is a Recurrent Neural Network (RNN) architecture with memory blocks in the hidden layers. The memory blocks have multiplicative gates that allow storing and accessing information over long periods. In this way, the vanishing gradient problem of the RNN is mitigated in the LSTM, since the gradient information is preserved over time. For this reason, LSTMs have been successfully applied to address real-world sequential and time-series problems [15]. The inputs consist of both the current sample and the previous observed sample, such that output at time step $t$-1 affects the output at time step $t$. Each neuron has a feedback loop that returns the current output as an input for the next step. This structure makes LSTMs an effective tool for prediction, especially in those cases where there is no previous knowledge about the extent of the time dependencies.

The inputs of our LSTM-based congestion predictor are denoted as a sample vector with the number of ECE-marked packets arriving at time intervals of 100 ms. This value corresponds to the typical assumption for the Round-Trip Time (RTT) in IP networks. Additionally, we rearrange that vector as an input matrix **X** corresponding to ten time steps and an output vector **y** of one time step, such that:

$$\mathbf{X} = \begin{bmatrix} x_{t_0} & x_{t_1} & \cdots & x_{t_9} \\ x_{t_1} & x_{t_2} & \cdots & x_{t_{10}} \\ \vdots & \vdots & \ddots & \vdots \\ x_{t_{N-10}} & x_{t_{N-9}} & \cdots & x_{t_{N-1}} \end{bmatrix}, \quad \mathbf{y} = \begin{bmatrix} x_{t_{10}} \\ x_{t_{11}} \\ \vdots \\ x_{t_N} \end{bmatrix} \quad (1)$$

where $x_{t_i}$ is the quantity of ECE-marked packets in the time interval $i$ and $N$ is the total number of samples. The rationale behind rearranging the samples in ten time steps is to improve the performance of the predictive model by having additional context. In this way, the estimation of arriving ECE-marked packets contemplates more prior observations.

For the design and training of the LSTM, we assume that the data are gathered in a ten-minute period, which is reasonable due to the dynamics of Internet networks. Consequently, there would be a dataset with 6000 samples, corresponding to the number of intervals of 100 ms in ten minutes. In addition, we consider an LSTM with three hidden layers and we use the approximation formula proposed in [16] to determine the number of neurons per layer, as follows:

$$N_n = (N_{in} + \sqrt{N})/L \quad (2)$$

where $N_{in}$ is the number of inputs, $N$ is the number of samples, and $L$ is the quantity of hidden layers. Then, $N_n \approx 30$ neurons per hidden layer. Finally, we take into account a dropout regularization of 20%, so that the model does not overfit and yields more generalized weights after training.

*B. Q-learning based AQM Parameter Tuner*

In general, the parameters of AQM algorithms are set to values that yield a reasonable performance for the typical network conditions. However, AQM mechanisms are expected to allow parameters adjustment depending on the specific characteristics of a network and their interactions with other network tasks over time [17]. Consequently, we embrace the idea of adjusting AQM parameters according to the network's changing circumstances, so that the performance is dynamically improved, as well. Nevertheless, the achievement of this goal can end up in a very complex job. For this reason, we propose a mechanism that adaptively tunes the parameters of the AQM in use as an RL-aided decision process.

We model the dynamic AQM parameter-tuning problem as a Markov Decision Process (MDP). Previous works have successfully modeled complex decision-making problems in networks through MDPs, [18]. For this intelligent method, the decision process is based on the inferred rest-of-path congestion, *i.e.* the output of our congestion predictor described in Section II-A. In this way, we define the states $S$ as a set of discrete levels of congestion that the flows will be likely to experience along the path, the set of actions $A$ comprises specific values of the target parameter, and the reward $R$ depends on the power function of the connection, which is defined as the throughput-to-RTT ratio. In our environment, the edge router acts as the agent that makes the decisions and, therefore, no extra intelligence is needed at the end devices. The idea behind using the predicted rest-of-path congestion is to proactively tune the AQM at the edge router. Consequently, our method can adjust the target parameter so that more packets are dropped instead of being marked, as they will be likely dropped ahead. On the other hand, if low congestion is forecasted ahead, the AQM will mark more packets based only on its own experienced congestion.

Nevertheless, finding the appropriate target for the balance between dropping/marking packets is a non-trivial problem and that is why we use RL. In other words, we model our problem as an MDP with the objective of finding an optimal behavior that maximizes the throughput-to-RTT ratio. To do so, we utilize the Q-learning algorithm [19], which defines a function $Q(S,A)$ representing the quality of a certain action in a given state and that is defined by:

$$Q(S,A) := Q(S,A) + \alpha \left[R + \gamma \max_a Q(S',a) - Q(S,A)\right] \quad (3)$$

where $a \in A$, $\alpha \in [0,1]$ is the learning rate, and the discount factor $\gamma \in [0,1]$ describes the preference of the agent for current rewards over future rewards. This equation characterizes the maximum future reward of present state and action in terms of immediate reward and maximum future reward for the next state $S'$. In this manner, the Q-learning algorithm iteratively approximates the function $Q(S,A)$.

More specifically, we model our AQM parameter tuner considering the current states as the observed levels of congestion, *i.e.* the ECE-marked packets arriving at the router in direction B, and the rest-of-path congestion prediction in direction A as possible next states. Both current and next states are discretized to delimit the complexity of the environment. On the other hand, the actions are a set of predefined values for the target parameter of the specific AQM in use.

III. EVALUATION METHODOLOGY AND RESULTS

In this section, we provide the details about the experimentation setup for the evaluation of our proposed solution. We first explain the preliminary experiments conducted to show the feasibility of our method as a whole, by studying the basis of each component separately. Later, we evaluate the performance of our intelligent AQM scheme comparing its operation to the behavior of conventional AQM. For our experimentation, we use the Mininet network emulator and the queue disciplines available in the Linux kernel. In this way, we validate the potential deployment of our solution in real network scenarios.

*A. Effects of Tuning AQM Parameters*

With respect to the AQM parameter tuner, in this work we evaluate our proposal using CoDel [3] and FQ-CoDel [5]. Therefore, the target parameter to tune is the acceptable minimum standing/persistent queue delay. To show the influence of changing the target parameter in both RTT and throughput metrics, we conducted some preliminary experiments by implementing a topology like the one depicted in Fig. 1. In the emulation scenario, the edge router on the left (R1) performs the AQM control and has 20 hosts, *i.e.* hosts B, connected to it. On the other side, 20 hosts connect to the right edge router (R2): these are hosts A. There are also a pair of monitor hosts, and one of them actively logs the measured RTT (mRTT) and throughput by means of sending probe packets to the other one. Note that for this experiment we consider a propagation delay of 20 ms and a bandwidth of 200 Mbps between hosts B and R1. Conversely, there is no propagation delay from R1 to R2 and, to emulate the path bottleneck, the link between the two routers has a bandwidth of 20 Mbps. The links between R2 and the hosts A have a bandwidth of 100 Mbps and no propagation delay. In addition, all hosts are ECN-enabled and each pair of hosts AB generates TCP traffic, mainly in direction A. In this work, we conduct our experimentation only with CUBIC, the default TCP congestion control in Linux.

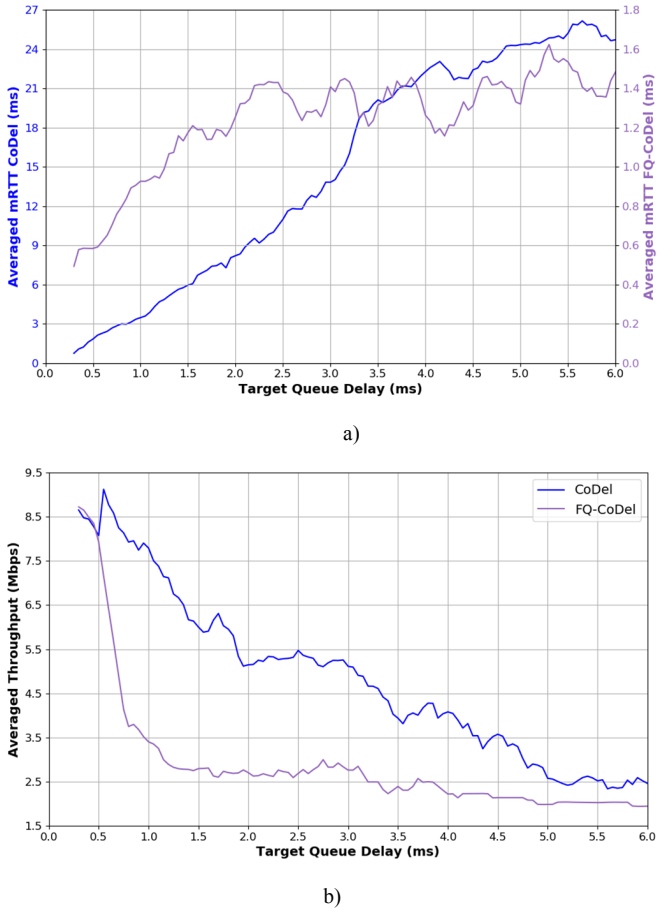

Fig. 2. Effects of varying the target parameter in CoDel and FQ-CoDel algorithms on: a) Averaged mRTT. b) Averaged throughput.

The experiment consists of modifying the target and interval parameters of CoDel and FQ-CoDel in R1, while data are constantly and simultaneously transferred from the hosts B to hosts A. The interval parameter ensures that the measured minimum delay does not become too old and, typically, the target delay is 5% of this interval. Therefore, we set CoDel and FQ-CoDel in R1 with target values from 50 µs to 6 ms and intervals from 1 ms to 120 ms, respectively. We left the other parameters as default, except the hard limit on the queue size, which we set to 1000 packets. This a configurable parameter set by the system administrator. Fig. 2 shows the resulting average mRTT and throughput for both queue disciplines in this experiment. Note that Fig. 2a has two different scales for the y-axis, since the mRTT is significantly longer for CoDel. As can be seen, although the target parameter of these AQM algorithms is meant to operate unchangeably, there is a noticeable effect when the target parameter varies. The lower the target queue delay, the more dropped packets, since not all packets can be ECN-marked when the router experiences congestion. Consequently, RTT is low and throughput is high when low target delay is configured, Fig. 2b. In other words, as the target parameter increases, the AQM mechanism produces more ECN-marked packets and drops less. This is consistent with our solution formulation explained in Section II-B.

*B. Transferring the Predictor Model*

As an initial training and test for our congestion predictor, we use the data from a backbone Internet link of an ISP collected by the Center for Applied Internet Data Analysis (CAIDA). The CAIDA's monitors collect packet headers at large peering points and a wide variety of research projects has used its anonymized traces [20]. Specifically, we use the data from the collection monitor that is connected to an OC192 backbone link (9953 Mbps) of a Tier 1 ISP, between New York, US, and Sao Paulo, Brazil. We use this dataset as valid data for an edge router, according to previous works cited at CAIDA's website, in which those data have been used similarly. In particular, we chose to analyze the data from December 20, 2018.

We perform the pre-training for the congestion predictor with data containing ECE-marked packets sent from New York to Sao Paulo, as we found that there are more ECE-marked packets in direction B than in direction A. According to the assumptions explained in Section II-A, we use the trace data in the ten-minute period with the highest number of ECE-marked packets that are not part of the ECN negotiation, that is from 8:00 to 8:10 EST. The traces show that, in this period, there were 402 different source IPv4 addresses sending ECE-marked packets to 315 destination hosts. We split the dataset into a training subset, corresponding to 80% of samples, and a test subset with 20% of samples. After 100 epochs of training, we test the model by making predictions with samples from the normalized subsets. We obtain a Root Mean Squared Error (RMSE) score of 0.08 and a Mean Absolute Error (MAE) score of 0.04 for the test subset. Similarly, we get an RMSE of 0.07 and a MAE of 0.03 for the training subset. Fig. 3 shows the actual normalized number of ECE-marked packets arriving at the router in direction B and the prediction over the test subset. As can be seen, the white spaces in the graph mean consecutive time intervals with no congestion, *i.e.* no ECE-marked packets at the router. On the other hand, the spikes depict time intervals in which congestion was experienced. Note that rather than predicting an accurate number of ECE-marked packets that will arrive, we model the predictor to estimate whether there will be significant congestion within the next time interval. In this way, Fig. 3a illustrates how the resulting prediction captures the tendency of the levels of congestion ahead.

Hence, we use the pre-trained LSTM model to accelerate the congestion estimation in our method. As the network conditions change, our method updates the predictor by re-training it with new data. However, this re-training process is much faster, as the LSTM updates in just one epoch, which takes about four seconds in our emulation environment. To see how the pre-trained congestion predictor behaves in a new environment, we run an experiment with the topology described in Section III-A. Moreover, to stress the network and make it more stochastic, we set random values of bandwidth and propagation delays on the links between hosts and routers. Likewise, each host B starts its transmission at a random time. The link bandwidth between R1 and R2 is the only non-random value fixed at 10 Mbps. Also, FQ-CoDel is the AQM method in this experiment with its default target delay and interval values, which are 5 ms and 100 ms, respectively.

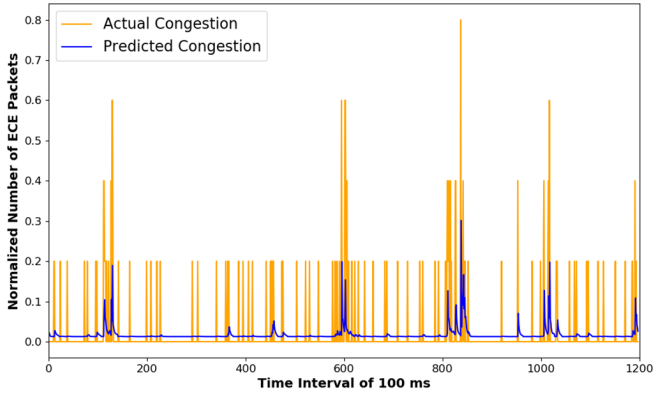

a)

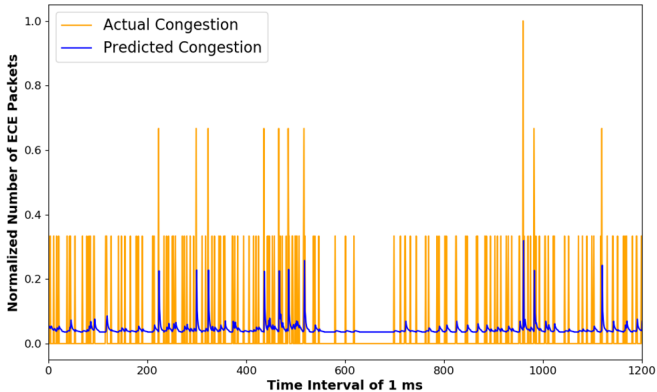

b)

Fig. 3. Actual and predicted congestion obtained after: a) Pre-training over 100 epochs, using the CAIDA's dataset. b) Re-training in one epoch, based on network emulation data.

In relation to the re-train process, we update the model with data gathered in six seconds. As we designed the congestion predictor for 6000 intervals (see Section II-A), we need to reduce the value of each time interval for the updates. Therefore, in this case, we re-train the LSTM with data in time intervals of 1 ms. After one update, the obtained values of RMSE are 0.09 and 0.13 for training and test subsets, respectively. In the same way, the resulting values of MAE are 0.04 for training and 0.06 for test. These scores show that our model can make predictions in the new network with a significant approximation and without the need for training the model from scratch. Fig. 3b depicts the congestion prediction results in the described network. Note that we scale by two times the graph corresponding to the prediction, *i.e.* the blue plot, for clarity of the comparison. Again, rather than the exact number of ECE-marked packets, we want to predict the congestion level trends ahead.

### C. Performance Evaluation of the Intelligent AQM

In this subsection we elaborate more about the experiments that we conducted to show the job of our proposed method as a whole. In Section II-B, we briefly described how the congestion predictor integrates with the AQM parameter tuner. We evaluate the MDP for this problem considering 100 levels of congestion as current or next states. The observed congestion corresponds to the current state and the predicted congestion is the next state. To determine their levels, we keep the maximum observed and predicted values as reference for the discretization. We also delimit the actions to 100 values, which in this case are the target delay of CoDel and FQ-CoDel. In this way, the possible actions are a set of values from 50 µs to 5 ms in steps of 50 µs. As we explained in Section III-A, we modify two parameters at the same time: the target delay and the interval. Thus, the experiments are more consistent, as these two parameters are tightly related. Again, the hard limit buffer size is set to 1000 packets and the TCP congestion control is CUBIC. The starting values for the target and the interval parameters are the default ones in the Linux kernel: 5 ms and 100 ms, respectively. For this evaluation, R1 performs the intelligent AQM while R2 needs only to be configured as ECN-enabled or as a regular router that does not wipe CE-marked IP packets.

Fig. 4 shows the results comparison when our intelligent method is applied to CoDel and FQ-CoDel, in terms of the cumulative power function. Note that these AQM schemes have static target parameters set to their default values when no intelligence is dynamically adapting them. As any other RL-based solution, the basic idea is to have an agent, *i.e.* the edge router in our problem, making decisions and getting feedback from the environment to calculate the rewards. To achieve so, we constantly capture the ECE-marked packets arriving at the router in direction B. Every second, the agent predicts the congestion of the rest-of-path in direction A. As the agent does not know what action to take at the beginning, there is an initial stage of exploration, which depends on the parameter $\varepsilon$. The value of this parameter determines if the Q-learning algorithm prefers to explore rather than exploit the historical data. In our experiments, we set $\varepsilon = 0.5$ so that the algorithm does not explore too greedily. After taking an action, either by randomly exploring or by extracting Q-values, the monitoring hosts measure the mRTT and throughput with active probes. We use these measures to calculate the power of the connection, which is our reward function. Once the rewards are known, the algorithm updates the Q-values by applying (3). Instead of updating the Q-values iteratively with a matrix containing predefined rewards, we train the model in an online manner by getting the feedback from the network. This could have the disadvantage of a poor behavior at the beginning, but the results show that the tuning improves over the time. We also point out that we implemented fixed values for the rest of the parameters of the Q-learning algorithm during the experiment, that is $\gamma = 0.8$ and $\alpha = 0.5$.

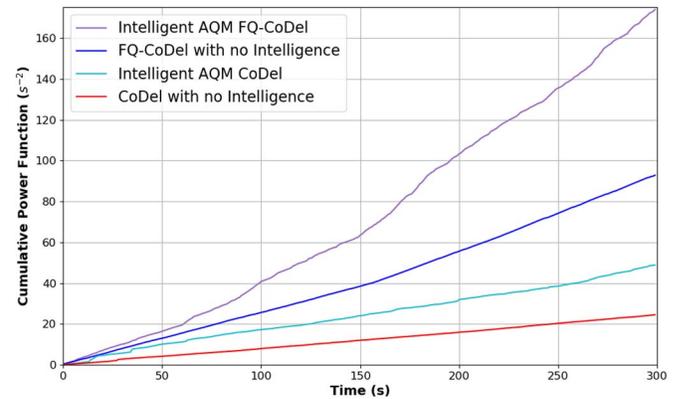

Fig. 4. Cumulative power of the connection measured during the experiments in the emulation environment. The intelligent method is applied to CoDel and FQ-CoDel. All schemes utilize ECN.

TABLE I. BUFFER OCCUPANCY COMPARISON

|  | Intelligent AQM | | Non-Intelligent AQM | |
|---|---|---|---|---|
|  | *Average* | *Maximum* | *Average* | *Maximum* |
| FQ-CoDel | 1.60 % | 2.70 % | 2.09 % | 2.80 % |
| CoDel | 0.91 % | 2.30 % | 1.58 % | 2.90 % |

Another point to consider is the performance of our method in terms of the buffer occupancy at the router. Based on the statistics obtained from the Linux Traffic Control utility, we compare the percentage of buffer occupancy for each experiment in Table I. Note that we take into account the set hard limit buffer size for the percentage calculation. In other words, the buffer occupancy would be 100% if the queue had 1000 packets at a specific instant. As can be seen, the buffer occupancy is lower when R1 employs our intelligent AQM, thanks to the balance between dropped/marked packets that the algorithm achieves over the time. Finally, we want to mention that the Python code of the experiments described in this subsection is publicly available at [21]. We intent to make our contribution accessible to researchers and developers who are actively working on congestion-related problems of the Internet. Please cite this paper if you use any posted script for your works.

IV. CONCLUSIONS

In this work, we showed how the appropriate tuning of AQM parameters can improve the RTT and throughput of TCP connections in a dynamic IP network. Additionally, we showed that it is possible to take advantage of the ECN mechanism to predict congestion on the rest-of-path. We modeled a congestion predictor based on an LSTM, which we pre-trained with data of an unknown network topology. We exposed how to transfer the predictor model to a new network and get good estimates with a rapid re-training. Also, we described a solution for the decision-making problem on the parameters that an AQM scheme should have according to the networks' conditions. We demonstrated that this can be achieved by modeling the problem as an MDP and finding pair values of state-action through the Q-learning algorithm. Although we employed the power function of the connection as the reward function, our method can work with other rewards depending on the applications or the TCP connection variable to be optimized. As a future work, we plan to test our proposed method with different TCP congestion control mechanisms, as well as more AQM algorithms. Finally, we point out that, although our experiments included only two AQM schemes with queue delay as the target parameter, the proposed intelligent method could be easily adapted to other schemes with different target parameters such as the queue size.